\documentclass[twocolumn,prl,showpacs,floatfix,preprintnumbers]{revtex4-1}
\usepackage{graphicx}
\usepackage{amsmath}
\usepackage{amssymb}
\usepackage{amsfonts}
\usepackage{upgreek}
\usepackage{color}
\usepackage{comment}
\usepackage{rotating}
\usepackage{mathdots}

\newcommand{\kb}{\mathbf{k}}
\newcommand{\kr}{\mathrm{k}}
\newcommand{\rb}{\mathbf{r}}

\newcommand{\kbz}{k_{\rm B}}

\begin{document}


\title{Umklapp Superradiance from a Collisionless Quantum Degenerate Fermi Gas}

\author{Francesco Piazza$^1$}
\email{francesco.piazza@ph.tum.de}
\author{Philipp Strack$^2$}
\affiliation{$^{1}$Physik Department, Technische Universit\"at M\"unchen, 85747 Garching, Germany}
\affiliation{$^{2}$ Department of Physics, Harvard University, Cambridge MA 02138}

\date{\today}

\begin{abstract}

The quantum dynamics of the electromagnetic light mode of an optical cavity 
filled with a coherently driven Fermi gas of ultracold atoms strongly depends on geometry of the Fermi surface.
Superradiant light generation and self-organization of the atoms can 
be achieved at low pumping threshold due to resonant atom-photon Umklapp processes, 
where the fermions are scattered from one side of the Fermi surface to the other by 
exchanging photon momenta. The cavity spectrum exhibits 
sidebands, that, despite strong atom-light coupling and cavity decay, 
retain narrow linewidth, due to absorptionless transparency windows outside the 
atomic particle-hole continuum and the suppression of inhomogeneous broadening and thermal fluctuations 
in the collisionless 
Fermi gas.

\end{abstract}

\pacs{PACS numbers}

\maketitle


{\it Introduction --} 
Lasing \cite{haken85,siegman86} and superradiance phenomena are currently enjoying a renaissance 
as research topics in atomic physics. Recent advances in quantum optical experiments with ultracold atoms 
enable the exploration of a new regime at ultralow temperatures in which quantum effects of both, 
the light and the atomic matter field become important, atoms and confined photon fields are strongly coupled, 
and the photon field is actually dynamical, playing a much more active role than in optical lattice experiments.

In many-body cavity quantum electrodynamics, where many ultracold 
atoms are placed in an optical resonator \cite{cavity_rmp},
the role of the photon is dual: first, it mediates 
an interaction between the atoms producing new phases of matter, 
whose dynamics, in turn, backacts on the photon field itself. Second, the photon output serves as a 
non-invasive probe of optical properties such as refraction and absorption of the underlying atomic medium.

An important question is in how far favorable coherence 
properties of the atomic medium can be 
imprinted onto the light field. To that end, a recent string of experiments has achieved 
superradiance \cite{dicke54} in which the $N$ atoms {\it collectively} interact with the light field:
Refs.~\onlinecite{inouye_1999,kuga_2005} with Raman photons, 
Refs.~\onlinecite{vuletic_2003,courteille_2007,barrett_2012,eth_2010,eth_jumps,eth_soft,eth_non_eq} 
with momentum recoil in thermal and condensed Bose gases, and Refs.~\onlinecite{weitz_2010,weitz_2013}
with photon gases in optical cavities.
In related superradiant lasers \cite{gross82,yamamoto99, haake_1993,meiser_2009,bohnet_2012}, 
the emitted intensity can be amplified $\sim N^2$ while there is substantial  $\sim1/N^2$ linewidth narrowing, with 
potential technological applications for precision spectroscopy and quantum metrology. 



\begin{figure}[t]	
\vspace{1mm}
\includegraphics[width=85mm]{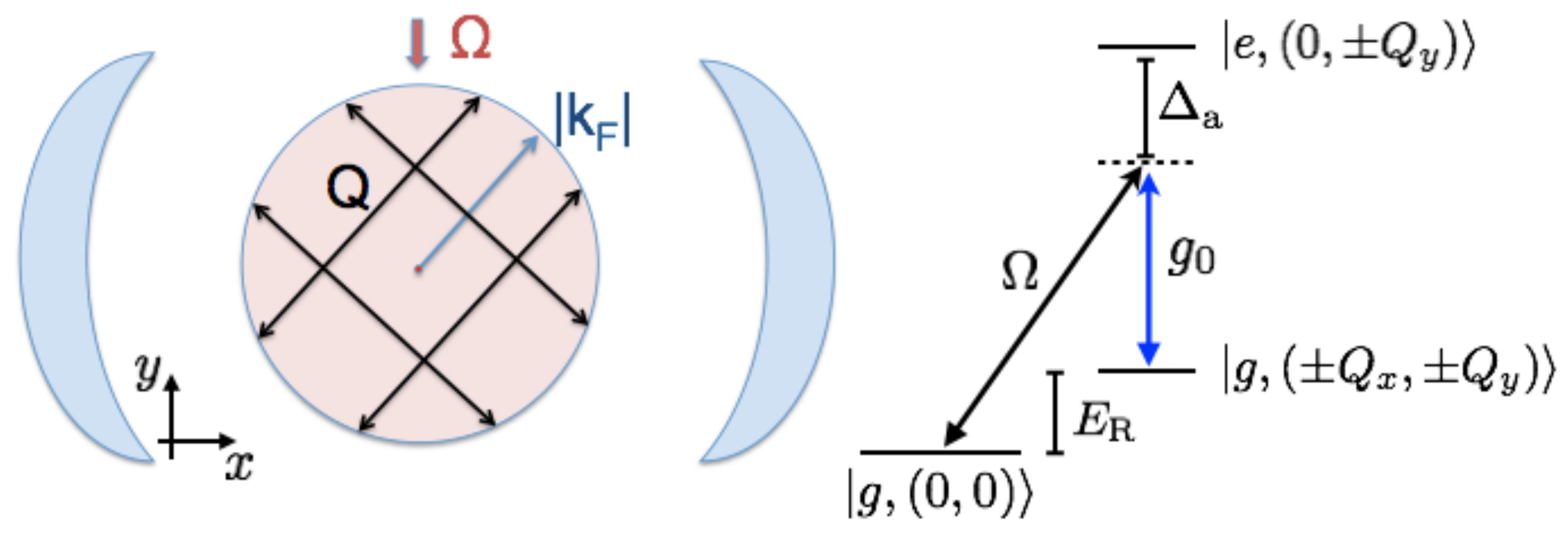}
\caption{Left: Fermi sphere (red circle) of the ultracold, atomic Fermi gas inside the mirrors of an optical cavity (curved, blue) in combined position-momentum representation. $|\mathbf{k}_{\text{F}}|$ is the Fermi momentum and $\mathbf{Q}$ is the superposition of the cavity momentum and the momentum of coherent drive laser with amplitude $\Omega$. 
Right: level scheme employed for the superradiant self-organization discussed here, with single-photon atom-cavity Rabi coupling $g_0$. Generalizing of the level structure above to an incoherent repumping scheme should enable 
superradiant  Umklapp lasing \cite{haake_1993,bohnet_2012,meiser_2009}. }
\label{fig:cavity_sketch}
\end{figure}

The purpose of this paper is to pin down the consequences of the Fermi surface in many-fermion cavity quantum 
electrodynamics not available with the previously discussed 
bosons or (effective) spins
\cite{
lieb_1973,
hioe_1973,
desalvo_1994,
ritsch_2002,
emary03,
emary_2004,
domokos_2005,
domokos_2006,
vukics_2007,
carmichael_2007,
maschler_2008,
domokos_2008,
larson_2009,
domokos_2010,
sarang_2010,
morigi_2010,
simons_2010,
simons10,
domokos_2011,
nagy11,
oztop12,
dallatorre_2012,
bhaseen_2012,
chang_2013,
griesser_2013,
hofstetter13,
oeztop13}.
Motivated by near-time experimental prospects to study superradiant phenomena with 
fermionic atoms in a transversally driven optical cavity (sketched in Fig.~\ref{fig:cavity_sketch}), 
we here provide a computation of the steady-state phase diagrams for one- and two-dimensional 
confinement of the Fermi gas as well as the cavity spectrum for this system. Single-spin fermions are appealing
as a coherent optical medium since frequency shifts from collisions are strongly suppressed, a feature also exploited in optical clocks \cite{hazlett13,katori08,ye11}. Optomechanics with fermions 
was considered previously in 
Refs.~\onlinecite{larson_2008,kanamoto_2010}, and glassy fermions in multi-mode cavities 
were discussed in Ref.~\onlinecite{mueller_2013}.

{\it Key results --} 
The Fermi gas is generically closer to superradiance threshold due to Umklapp scattering events between points on the Fermi surface transferring the two-photon momentum $\mathbf{Q}$ at no energy cost. The density and the confinement dimensionality of the Fermi gas drastically affect available phase space volume for low energy Umklapp processes. Atomic self-organization, concomitant with superradiance, occurs here 
as a dynamical Peierls instability without 
a preformed, conservative lattice potential parallel to the cavity axis. In $d=1$, perfect nesting between $\mathbf{Q}$ and $\mathbf{k}_{\text{F}}$ strongly reduces the critical pump strength towards a fermionic superradiant state, 
which can become fully insulating. In d=2, cavity-mediated Fermi surface reconstruction leads to 
"pockets" of gapless excitations similar to magnetic metals \cite{vojta12}.

\begin{figure}[t]	
\includegraphics[width=85mm]{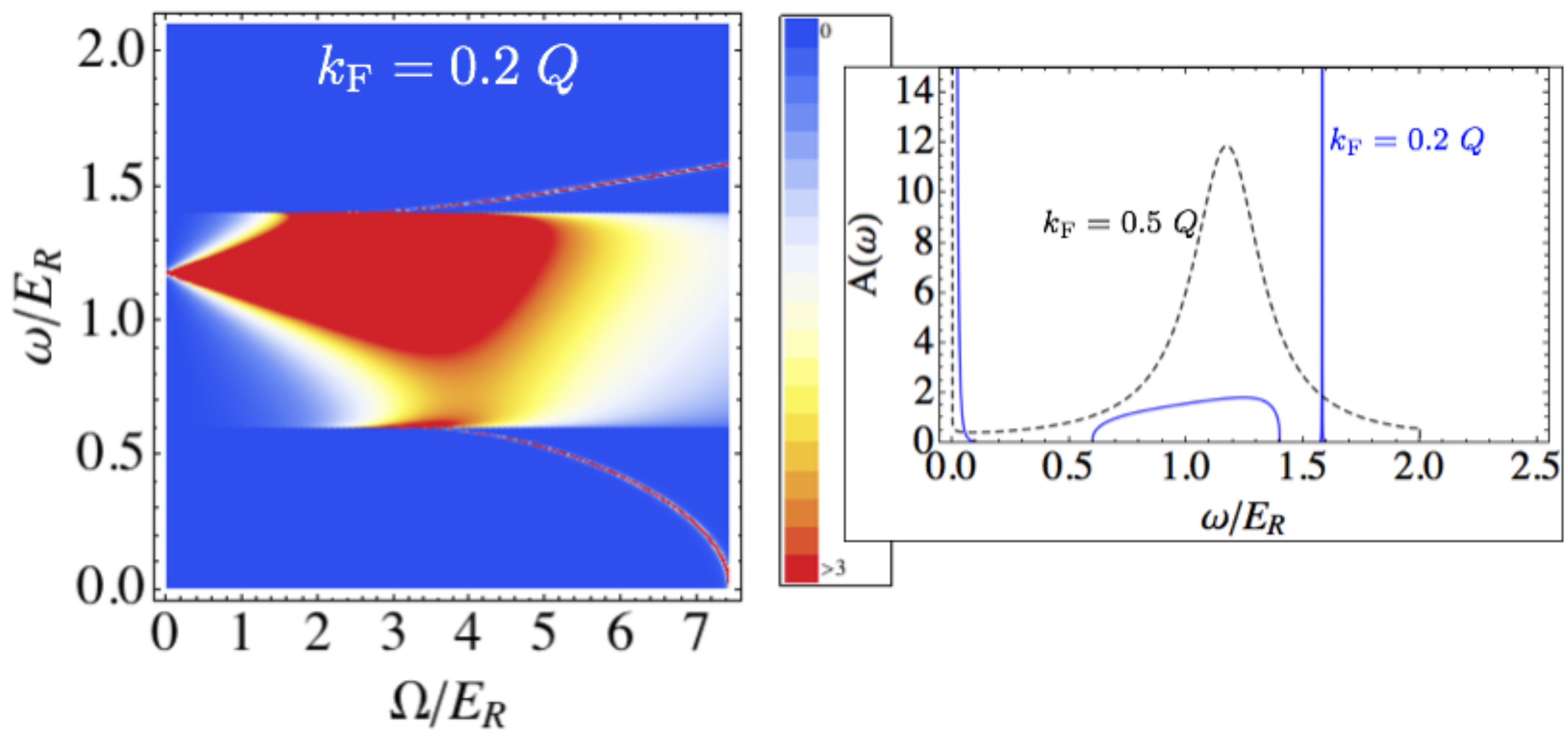}\\[-3mm]
\caption{Cavity photon spectrum in one dimension. 
Left panel: spectral function Eq.~(\ref{eq:inside}) in color scale. 
The narrow features are actually sharp $\delta$-function peaks, the lower of which moves toward $\omega=0$ for $\Omega\rightarrow \Omega_{\rm D}$. Right panel: spectral function at fixed coupling close to threshold, $\Omega =0.98\Omega_{\rm D}$, for two different densities.
Black-dashed curve: perfectly nested case in which the fermionic particle-hole continuum reaches 
down to zero frequency and $\Omega_{\rm D}=0$. 
Parameters are $\Delta_{\rm c}=-1.2\;E_{\rm R},\;Ng_0^2/\Delta_{\rm a}=-0.05\;E_{\rm R},\;g_0/\Delta_{\rm a}=-0.1\;E_{\rm R}, T=0$.
Within the particle-hole continuum there is ``broadband'' emission for 
a range of frequencies, outside of the particle-hole continuum, the linewidth remains narrow.}
\label{fig:spectral_density}
\end{figure}

\begin{figure}[t]	
\includegraphics[width=50mm]{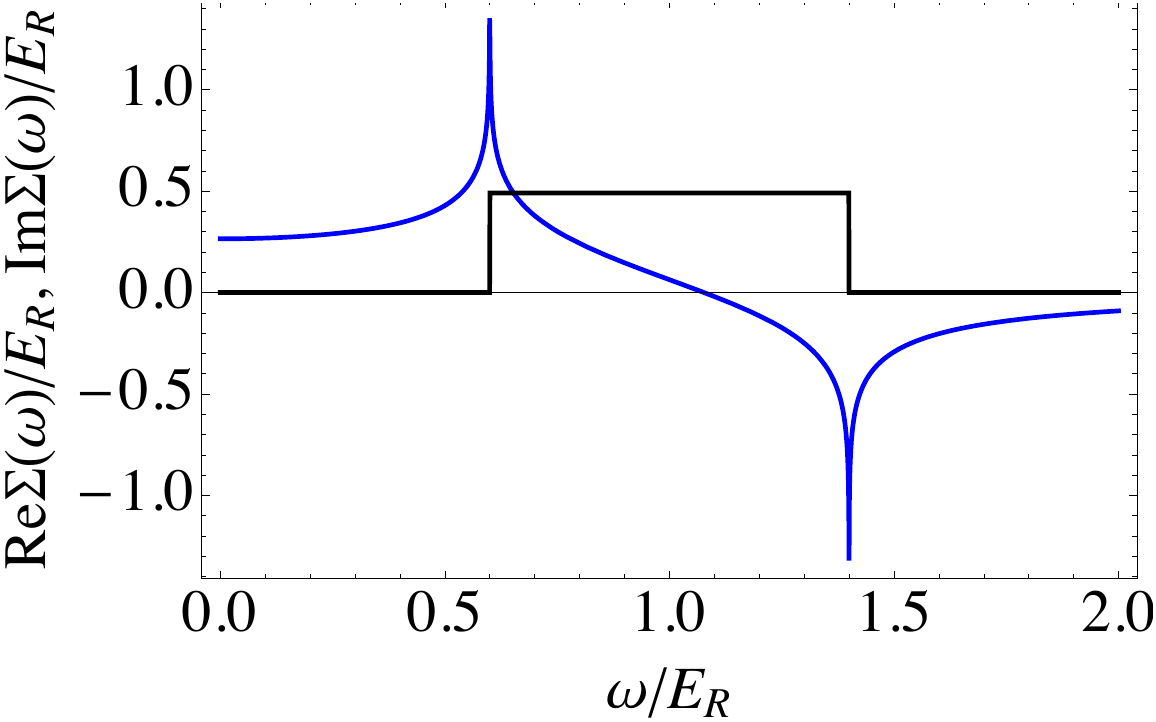}\\[-3mm]
\caption{
Absorption (imaginary part of the particle-hole continuum, black line) and 
refractive properties (real part of the particle-hole continuum, blue line) of the 
coherent atomic Fermi medium in one dimension at $\Omega=0.6\Omega_{\rm D}$.
The lower sideband in Fig.~\ref{fig:spectral_density} appears in 
the absorptionless "transparency window" for $\omega/E_R \lesssim 0.6$, for which 
the imaginary part is identically zero since here $k_{\text F}=0.2Q$ is away from perfect nesting. At 
perfect nesting $Q  = 2 k_F$ the absorption reaches to $\omega = 0$. The optical properties 
of the atomic Fermi medium can be tuned by the coherent drive $\Omega$ 
highlighting connections to electromagnetically induced transparency \cite{eit_review}.}
\label{fig:eit}
\end{figure}

The cavity spectrum, instead of the usual broadened polaritonic peaks, shows a broad continuum with sharp edges plus two perfectly narrow sidebands shown in Fig.~\ref{fig:spectral_density}. Sufficiently close to the self-organization transition, 
the sidebands appear in "absorptionless" transparency windows (below $\omega/E_R \lesssim 0.6$) and 
as such they remain perfectly sharp. The underlying absorption and refractive 
properties of coherent atomic Fermi medium, shown in Fig.~\ref{fig:eit}, are determined by the imaginary and real part of the 
particle-hole continuum, respectively. Moreover, we expect the sidebands to be robust agains thermal noise, due to the absence of collisions in the single-spin Fermi gas and the associated reduced sensitivity to thermal fluctuations (especially in interacting Bose superfluids, order 
parameter fluctuations will strongly deplete the condensate fraction available for narrow recoil lasing); the far-detuning between the internal atomic transition and the cavity frequency also ensures that decay from the excited atomic state is surpressed.
Currently used Fabry-P\'erot cavities with MHz decay rates are in the bad cavity limit compared to the 
"slow motion" of alkali atoms with kHz kinetic energies, leading to a decoupling of the photon decay and the collective 
dynamics of the atoms \cite{meiser_2009,bohnet_2012,eth_soft,eth_non_eq}. Indeed, our results here are  
confirmed in a non-equilibrium calculation \cite{piazza14}.


%

{\it Set-up and formalism --} 
We consider $N$ spinless fermionic atoms with two internal electronic levels in the setup of Fig.~\ref{fig:cavity_sketch}. The quantized excitations of the coupled atoms plus driven cavity system will be 
described in terms of the field
operators $\hat{\psi}_{g/e}$ for the atoms in the internal ground or excited state
and the annihilation operator $\hat{a}$ for a cavity photon~\cite{maschler_2008}.
The atomic operators obey fermionic quantum statistics and 
fulfill the (anti-) commutation relation $\left\{\hat{\psi}(\mathbf{r}),\hat{\psi}^\dagger(\mathbf{r}')\right\}=
\delta_{\mathbf{r},\mathbf{r'}}$.
In a frame rotating with the frequency $\omega_p$ of the pump laser,
the Hamiltonian 
$
\hat{H}=\hat{H}_{\rm a}+\hat{H}_{\rm c}+\hat{H}_{\rm a/c}+\hat{H}_{\rm
  a/p}$ 
contains four terms: $\hat{H}_{\rm c}=-\Delta_{\rm c}\ \hat{a}^{\dag}\hat{a}^{}$ and 
\begin{align}
&\hat{H}_{\rm a}=-\int
d\mathbf{r}\left[\hat{\psi}_g^{\dag}(\mathbf{r})(\frac{\nabla^2}{2
    m})\hat{\psi}_g^{}(\mathbf{r})+\hat{\psi}_e^{\dag}(\mathbf{r})(\frac{\nabla^2}{2
    m}+\Delta_{\rm a})\hat{\psi}_e^{}(\mathbf{r})\right]\nonumber\\
&\hat{H}_{\rm a/c}=-i\ g_0\int d\mathbf{r} \hat{\psi}_g^{\dag}(\mathbf{r})\eta_{\rm c}(\mathbf{r})\hat{a}^{\dag}\hat{\psi}_e^{}(\mathbf{r})
+ {\rm h.c}\nonumber\\
&\hat{H}_{\rm a/p}=-i\ \Omega\int d\mathbf{r} \hat{\psi}_g^{\dag}(\mathbf{r})\eta_{\rm p}(\mathbf{r})\hat{\psi}_e^{}(\mathbf{r})+ {\rm h.c}
\label{hamiltonian}
\end{align}

Here, $\hat{H}_{\rm a}$ describes the kinetic energy of the atoms with mass $m$ moving around inside the cavity, 
with the excited state detuning between the pump and the atomic resonance $\Delta_a = \omega_p - \omega_c$. 
$\Delta_{\rm c}=\omega_{\rm p}-\omega_{\rm c}$ is the detuning between
the pump and the cavity mode and $\Omega$ is the pump Rabi frequency. 
We operate in the standard regime where the atoms couple to
only a single excitation mode of the electromagnetic field of the
cavity with single-photon Rabi coupling $g_0$. 
The functions $\eta_{\rm c}(\mathbf r)=\cos(\mathbf{Q}_{\rm c} \cdot \mathbf{r})$ and
$\eta_{\rm p}(\mathbf r)= \cos(\mathbf{Q}_{\rm p} \cdot \mathbf{r})$ (we choose $\mathbf{Q}_{\rm c(p)}=Q_{x(y)}\mathbf{\hat{x}}(\mathbf{\hat{y}})$ below), contain the
spatial structure of the mode functions of the (standing-wave) cavity light field and the pump laser, respectively. 

We extend our recently developed effective action formalism \cite{piazza13} to fermionic quantum fields and 
to situations with a spatially varying pump laser potential. We neglect the spontaneous emission from the excited
atomic level by assuming that the detuning $\Delta_{\rm a}$ be by far
the largest energy scale such that population of the excited level is suppressed. This allows us to adiabatically eliminate
the excited atomic level and derive an effective action for the low-lying levels coupled to the cavity 
(see Supplemental Material). The effective action is amendable to a 
saddle-point analysis at fixed density
$n_{\psi}=N/L^d$, as done in \cite{piazza13}. The corresponding mean-field
solution $\alpha=\langle \hat{a} \rangle$ becomes exact in the thermodynamic limit $N,L\to\infty, n_{\psi}=\text{const.}$,
yielding the phase diagram as a function of 
the pump strength $\Omega$,the fermion density
$n_\psi$, dimensionality $d$, and temperature $T$. 
\begin{figure}[t]	
\vspace{-1mm}
\includegraphics[width=80mm]{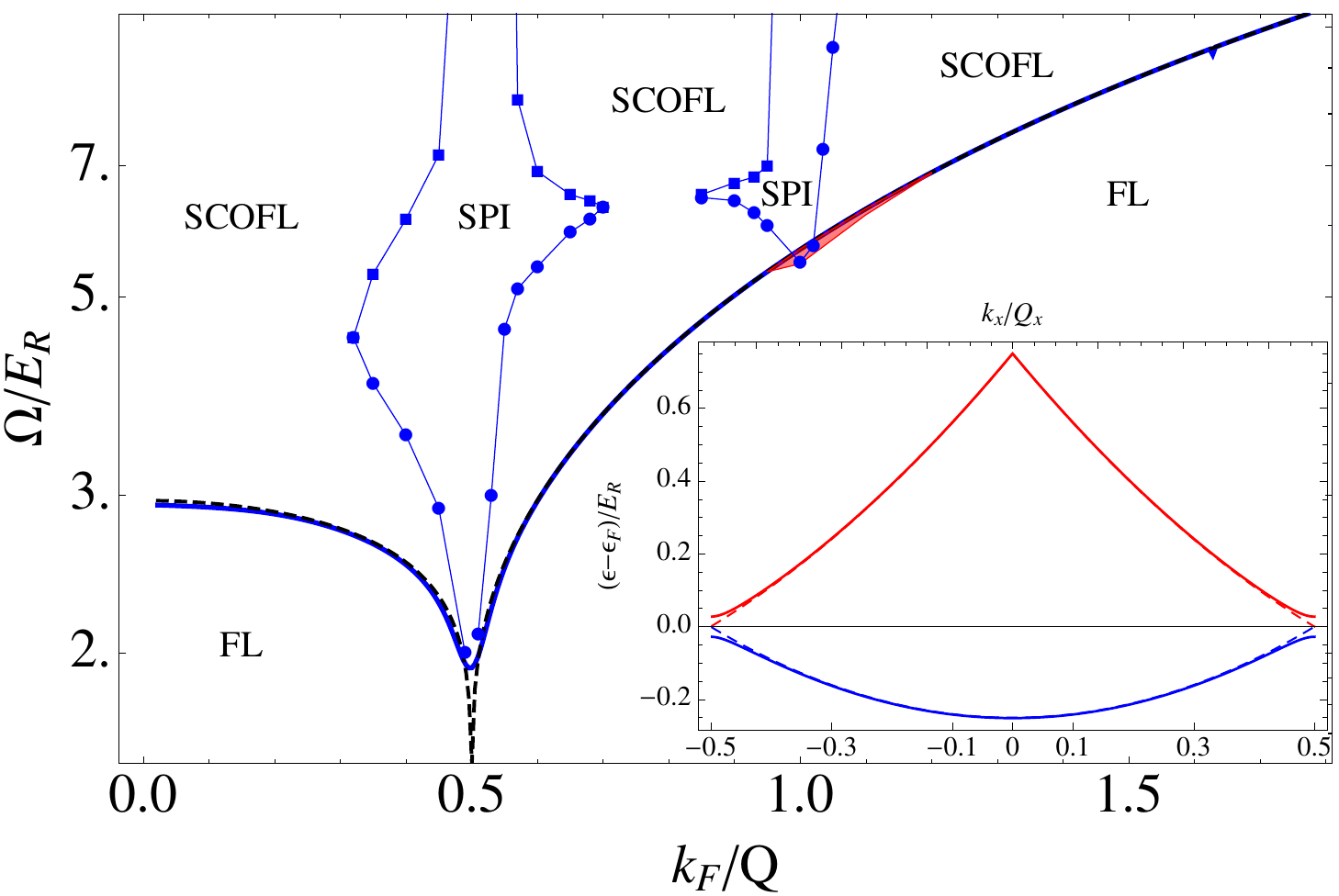}\\[-1mm]
\caption{Phase diagram for a one-dimensional Fermi gas in an optical
  cavity as a function of Fermi- ($k_F$) over cavity ($Q=\sqrt{2mE_{\rm R}}$) momentum versus 
  pump amplitude $\Omega$. Inset shows gap opening at the Fermi points. 
  Temperatures are $\kbz T=0$ (black-dashed line), $\kbz
  T=0.01E_{\rm R}$ (blue-solid line). At $Q = 2 k_F$ the system is perfectly nested and 
  Peierls reconstruction into a superradiant Peierls insulator (SPI) sets in at relatively small $\Omega$.
  Away from nesting the $Z_2$ charge symmetry breaking leads to a 
  superradiant charge-ordered Fermi liquid (SCOFL).
The red-shaded area extending from the second order
transition line indicates the hysteresis region preceding the
first-order phase transition from the Fermi liquid (FL) into the 
the SPI at $k_F/Q=1$; there the Fermi energy lies in the band 
between the second and third cavity generated bands. The remaining parameters are
$\Delta_{\rm c}=-0.2E_{\rm R}, Ng_0^2/\Delta_{\rm
  a}=-0.05E_{\rm R}, g_0/\Delta_{\rm a}=-0.1$.}
\label{fig:phase_diag_1d}
\end{figure}

{{\it Self-organization for 1d confinement -- }}
The phase diagram for the $d=1$ case, when the fermionic atoms are tightly confined in tubes 
parallel to the cavity axis so that the two-photon momentum transfer $\mathbf{Q}=\mathbf{Q}_{\rm c}+\mathbf{Q}_{\rm p}\simeq\mathbf{Q}_{\rm c}$, is shown
in Fig.~\ref{fig:phase_diag_1d}. The critical pump strength
$\Omega_{\rm D}$, above which the system is self-organised/superradiant,
strongly depends on the fermion density or, equivalently, on the 1D
Fermi momentum $k_{\text F}^{1\text D}=\pi n_{\psi}$. In particular,
we notice a strong suppression of $\Omega_{\rm D}$ when $k_{\text
  F}^{1\text D}\simeq Q/2$. This condition indeed implies
that a cavity photon can scatter an atom from the Fermi surface (for
$d=1$ Fermi points) at very low energy cost with a momentum transfer
$\mathbf{Q}$ which inverts the direction of the atomic motion (Umklapp scattering). 
For fermions in $d=1$, 
the system becomes unstable towards superradiance even at infinitesimal
pump strength for $T=0$ and $k_{\text
  F}^{1\text D}= Q/2$. The $T=0$ line in
Fig.~\ref{fig:phase_diag_1d} goes indeed to zero like
$1/\ln|1-Q/2k_{\text F}|^{-1}$, while as
soon as a small finite temperature (also potentially induced by cavity decay) is present $\Omega_{\rm D}$
stays finite. This is analogous to the Peierls instability present in
one-dimensional metals \cite{peierls_1955}, where it becomes
energetically favorable for the electrons to break the discrete
translational symmetry by doubling the lattice period so that the
Fermi points get gapped out, with the important difference
that here the cavity generated lattice does not reorganize but rather first appears due to
the instability. Ref.~\onlinecite{ulf13} is a related proposal to simulate Peierls 
  physics with hybrid ultracold atom and ion systems.
Our system becomes
insulating in the superradiant phase for nearly commensurate densities
$k_{\text F}^{1\text D}\simeq {\rm j}\ Q/2$ (with j integer) as is shown for 
${\rm j}=1,2$ in Fig.~\ref{fig:phase_diag_1d}. The superradiant Peierls insulating
regions are separated by crossover lines from regions where the system
shows superradiant charge order but is still metallic, since the Fermi
energy does not lie within the band gap. 
In addition to measurements of the cavity spectrum (see also below), 
the superradiant SPI and SCOFL phases could be distinguished by 
radio-frequency spectroscopy of the atomic cloud.
The FL-SCOFL transition is always continuous except from a region around
$k_{\text F}\simeq Q$ where the transition is first order. The
red-shaded area in Fig.~\ref{fig:phase_diag_1d} shows the 
region where the free energy has two local minima as a function of
the order parameter $\alpha$. This hysteresis region appears for $\Omega$ slightly lower
than $\Omega_{\rm D}$ and ends exactly at $\Omega_{\rm D}$, where the
free energy has only a single minimum at finite $\alpha$,
corresponding to the jump in the cavity occupation. Since the atoms couple to a single cavity mode
extending all over the cloud, there is no 
coexistence between the normal and superradiant phase despite hysteresis.

\begin{figure}[t]	
\vspace{-1mm}
\includegraphics[width=80mm]{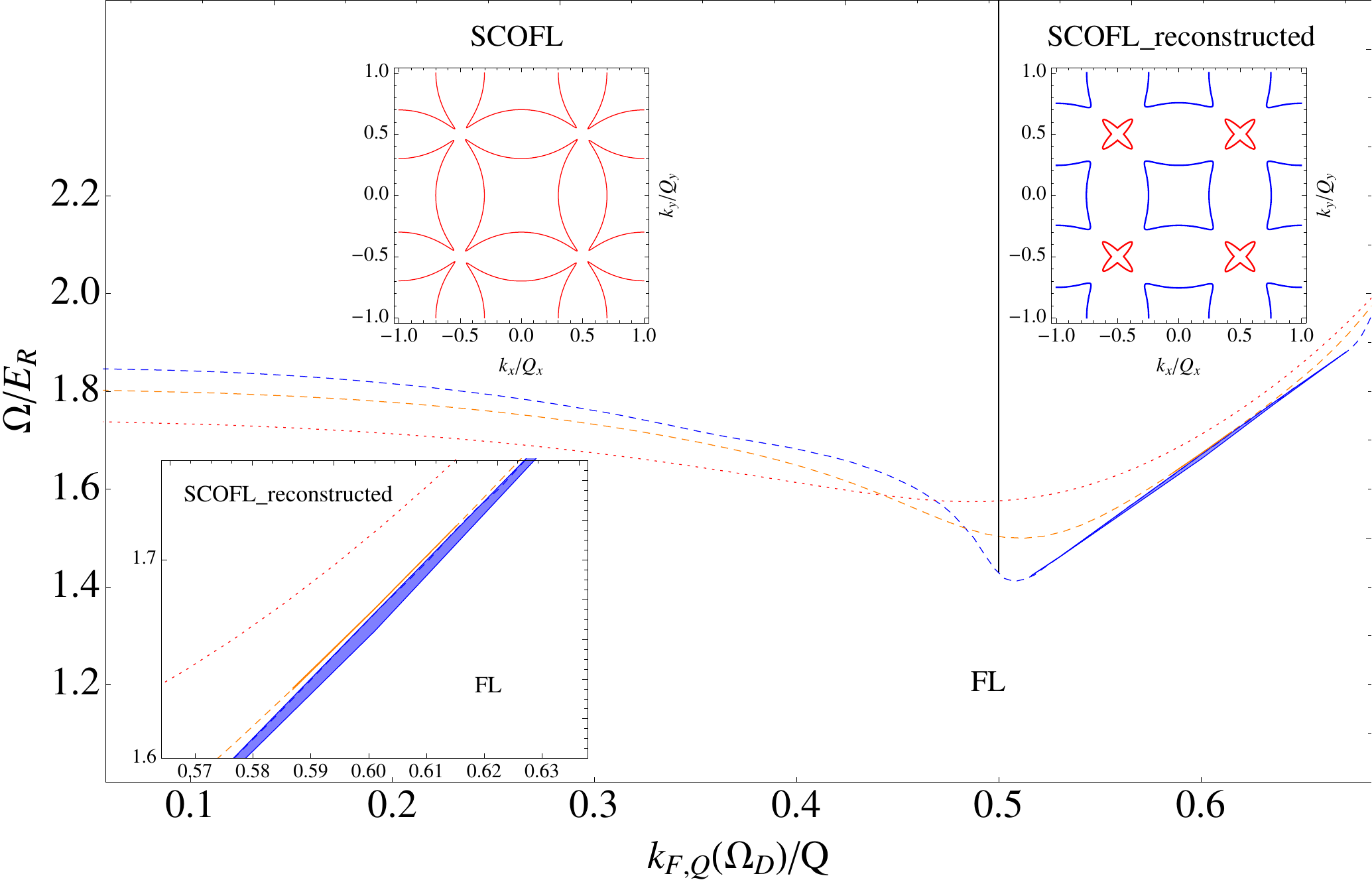}
\caption{Phase diagram for a two-dimensional Fermi gas in an optical
  cavity for $Q_x=Q_y=\sqrt{2mE_{\rm R}}$,
$\kbz T=0.01E_{\rm R}$ (blue-dashed line), $\kbz
  T=0.05E_{\rm R}$ (orange-dashed line), and $\kbz T=0.1E_{\rm R}$ (red-dotted line).
Lower inset: hysteresis region. 
The vertical line separates the two different superradiant regimes with topologically trivial Fermi surface and
reconstructed Fermi surface, as illustrated in the two upper
insets: left at $k_{\text F,\hat{\mathbf{Q}}}(\Omega_{\rm D})=0.49Q, \alpha=0.1$, right
$k_{\text F,\hat{\mathbf{Q}}}(\Omega_{\rm D})=0.53Q, \alpha=0.2$. 
Here the Fermi surface is shown in the repeated-zone scheme relative to the Brillouin zone
$\mathcal{B}=(-Q_x/2<\kr_x<Q_x/2, -Q_y/2<\kr_y< Q_y/2 )$. Purple lines delimit ``electron'' pockets with occupied
levels while blue lines delimit hole pockets with empty levels.
The vertical line separating the two superradiant phases is straight only close
to $\Omega_{\rm D}$.
The remaining parameters are as in Fig.~\ref{fig:phase_diag_1d}.
}
\label{fig:phase_diag_2d}
\end{figure}

{{\it Self-organization for 2d confinement -- }} In two dimensions the physics is richer since the spatial structure of
the pump laser cannot be neglected. This has two main effects: i) even
in the normal phase, by increasing the pump strength we deform the
Fermi surface of the atoms inside the pump lattice with vector
$\mathbf{Q}_p$, ii) the density wave in the superradiant phase has
momentum $\mathbf{Q}=\mathbf{Q}_c+\mathbf{Q}_p$, corresponding to a
chequerboard lattice with reciprocal vector $\mathbf{Q}$ whose length
is $Q=\sqrt{Q_x^2+Q_y^2}$ \cite{suppl}.
The phase diagram for different temperatures as a function of the
Fermi momentum along the $\mathbf{Q}$ direction (the relevant nesting direction) calculated at the critical pump strength $k_{\text F,\hat{\mathbf{Q}}}(\Omega_{\rm D})$ is presented
in Fig.~\ref{fig:phase_diag_2d}. As in the $d=1$ case, we observe a
suppression of $\Omega_{\rm D}$ for $k_{\text F,\hat{\mathbf{Q}}}(\Omega_{\rm D})\simeq Q/2$, marked by the vertical black-dashed line in
Fig.~\ref{fig:phase_diag_2d}.
The suppression of $\Omega_{\rm D}$ is much weaker as compared to
$d=1$ since perfect nesting is absent. Again, the minimum in
$\Omega_{\rm D}$ is at $k_{\text F,\hat{\mathbf{Q}}}(\Omega_{\rm D})\simeq Q/2$, where this time the Fermi momentum depends on the
pump strength due to the deformation of the Fermi surface
discussed at i). 
In Fig.~\ref{fig:phase_diag_2d}, the minimum is not exactly at $k_{\text F,\hat{\mathbf{Q}}}(\Omega_{\rm D})= Q/2$ since $T\neq 0$.
In addition, the self-organization transition in $d=2$ can correspond
to reconstruction of the Fermi surface for $k_{\text
  F,\hat{\mathbf{Q}}}(\Omega_{\rm D})> Q/2$, an example of which is
given in the right upper inset of Fig.~\ref{fig:phase_diag_2d}.
By entering the self-organized phase the atoms change from a simply
connected Fermi surface to one consisting of separated closed surfaces
delimiting zones with occupied and empty states. The blue square-shaped Fermi
surface belongs to the first chequerboard lattice band while the red star-shaped surface to the
second higher band. The condition $k_{\text F,\hat{\mathbf{Q}}}(\Omega_{\rm D})>Q/2$ implies that the Fermi surface has to intersect the first Bragg
plane of the reciprocal chequerboard lattice \cite{suppl}.
These phases with different Fermi surface topologies could be distinuguished by time-of-flight imaging of the atomic cloud.

 %

In contrast to the one-dimensional case, in $d=2$ the atomic medium is
therefore always metallic in the superradiant phase.
The order of the superradiant transition depends on $k_{\text F,\hat{\mathbf{Q}}}(\Omega_{\rm D})$ and temperature. As can be seen from
Fig.~\ref{fig:phase_diag_2d}, there is a large region where the
transition is first order and hysteresis is present (indicated by the
shaded area). The range of densities for which the transition is first order 
gets smaller with increasing temperature (see
lower inset in Fig.~\ref{fig:phase_diag_2d}).
In $d=2$, the Fermi momentum at $\Omega=0$ reads 
$|\mathbf{k}_{\text{F}}^{\text 2D}|=\sqrt{4 \pi n_\psi}$ and the cavity momentum is 
$Q_x = 2 \pi/\lambda$ with $\lambda$ an optical wavelength for example $\sim800$ nm.
For a degenerate Fermi gas at densities around $10^{14} - 10^{15}$ $m^{-2}$ the various regimes discussed here are thus experimentally accessible. 

{{\it Cavity spectrum for 1d-confinement -- }} 
We now turn to the cavity spectrum (shown above in Fig.~\ref{fig:spectral_density}), 
which is as well dramatically affected by the presence of a sharp Fermi surface at low $T$. The cavity spectral function:
\begin{align}
A(\omega)=
\frac{ -2 \left(\delta_c + \omega\right)^2 \text{Im}\Sigma(\omega)}
{\left(\delta_c^2 - \omega^2 + 2\delta_c \text{Re}\Sigma(\omega)\right)^2
+\left(2\delta_c \text{Im} \Sigma(\omega)\right)^2}\;,
\label{eq:inside}
\end{align}
with the shifted cavity detuning $\delta_{\rm c}=-\Delta_{\rm c}+Ng_0^2/2\Delta_{\rm a}$, describes how the spectral weight is distributed between different frequencies under a weak probe but, being normalized to $2\pi$, does not contain information about the intensity.  
The function $\Sigma(\omega) = -\frac{\lambda^2}{2n} \Pi_{\text{F}}(\omega, \mathbf{Q})$ (shown in Fig.~\ref{fig:eit}) describes how a photon is dressed by atomic fluctuations: its real part shifts the photon frequency while its imaginary part gives rise to broadening. Here $\lambda=\Omega g_0/\Delta_{\rm a}$.
We will focus on the zero temperature case
. Explicit formulae for the particle-hole polarization function $\Pi_{\text{F}}(\omega, \mathbf{Q})$ are given in the Supplemental Material. For energies within the particle-hole continuum, $(Q/k_{\text{F}})^2 - 2Q/k_{\text{F}} \leq \omega/\epsilon_{\text F} \leq (Q/k_{\text{F}})^2 + 2 Q/k_{\text{F}}$, the broadening of the photon 
spectrum is determined by a frequency independent constant 
\begin{align}
\text{Im}\Sigma(\omega)_{1d}=-\frac{\lambda^2 \pi k_{\text{F}}}{ 8 \epsilon_F Q}\;.
\label{eq:im_pi}
\end{align}
This broadening ("Landau damping") arises from real scattering events between photon and atom where momentum 
and energy are conserved $\omega=\epsilon_{k+Q}-\epsilon_k$.
Instead of normal-mode split polariton peaks in the cavity spectrum (with $A(\omega) $ being sum of two 
Lorentzians), Fig.~\ref{fig:spectral_density} exhibits a broad feature disappearing with a discontinuity 
for frequencies out of the particle-hole continuum, where Im$\Sigma(\omega)=0$.  
%
Moreover, when the particle-hole fluctuations in the atomic medium shift (through Re$\Sigma(\omega)$) the cavity 
frequency outside the particle-hole continuum, the cavity spectrum shows sharp sidebands 
%
$A(\omega) = \left(1+\delta_c \text{Re}\Sigma(\omega = E)\right)
\pi \delta(\omega - E)$\;.
%
The absence of damping in the low frequency range is due to the Pauli principle, which forbids to scatter a fermion into an occupied state, while in the high frequency range it is due to the sharpness of the Fermi surface, such that suddenly no fermions are available for scattering above a given threshold. 
In particular, the lower sideband is found
at the ``soft mode'' energy $E$, which close to the critical point is $E \simeq \sqrt{\delta_c^2 + 2 \delta_c \text{Re}\Sigma(\omega=0)}$ and goes zero at the critical pump strength $\Omega_{\rm D}$.
Close to perfect nesting $Q=2k_{\text F}$, see right panel of Fig.~\ref{fig:spectral_density}, 
the broad spectral feature reaches down to $\omega = 0$, leaving no space for a sharp soft mode.

{\it Summary --}
We considered the optical properties and self-organization of a coherently 
driven Fermi gas strongly coupled to the light field of an optical resonator.
The Fermi surface turned out to be the game changer leading to 
superradiance and self-organization at low pumping threshold as well 
as the appearance of narrow sidebands in the cavity spectrum. 
Generalizations of this work could lead to superradiant 
Umklapp lasers as well as new non-equilibrium 
phases of interacting fermions.

{\it Acknowledgments --} We are grateful to W. Zwerger for collaboration and guidance on related work, 
and to M. D. Lukin and H. Ritsch for insightful discussions and references on lasing. 
This work was supported by the Alexander Von Humboldt foundation, the DFG under grant Str 1176/1-1, by the 
NSF under Grant DMR-1103860, by the Templeton foundation, 
by the Center for Ultracold Atoms (CUA), and by the Multidisciplinary University Research Initiative (MURI).

\newpage

\begin{widetext}
\section{Supplemental Material}

\subsection{Photon-only action}
In order to obtain a photon-only effective action, we generalize our recently developed formalism \cite{piazza13} to Fermi statistics and also to include a spatially varying pump laser potential. We map the Hamiltonian given in the main text to an action using an infinite-dimensional atomic Bloch vector
%
$\Psi^{T}(\kb)=\Big(
\begin{array}{cccccccccc}
\psi_{k_x,k_y} &\psi_{k_x - Q_x,k_y} & \psi_{k_x + Q_x,k_y}&
\psi_{k_x ,k_y-Q_y} &
\psi_{k_x ,k_y+Q_y} & ...\end{array}\Big)$
%
 that exhaustively divides momentum space into cavity-generated 
``bands''. The latter are differentiated by multiples of cavity momentum $\mathbf{Q}_c = (Q_x,0)$ in $x$-direction and multiples of the pump 
laser momentum $\mathbf{Q}_p = (0,Q_y)$ in the $y$-direction transversal to the cavity axis. $Q_x$ and $Q_y$ need not 
be equal. The momentum sums are now restricted to the first Brillouin
zone $\kb\in \mathcal{B}$,  i. e. $\kb=(-Q_x/2<\kr_x<Q_x/2, -Q_y/2<\kr_y< Q_y/2 )$.

After first eliminating the atomic excited state adiabatically and then the atomic ground state exactly (since the Hamiltonian is quadratic)
we arrive at the photon-only action:
\begin{align}\label{S_eff}
&S_{\rm eff}[a^*\!,a]=\frac1\beta\sum_{\rm n}(-i\omega_{\rm
  n}-\Delta_{\rm c})|a_{\rm n}|^2- {\rm Tr}\ln\left[M_{\rm n,m}(\mathbf{k})\right]\ ,
\end{align}
with ${\rm Tr}=\sum_{\rm n}L^d\int_{\mathcal{B}}\frac{d\kb}{(2\pi)^d}\ {\rm tr}$, where
${\rm tr}$ is matrix trace in Nambu space over the formally infinite-dimensional matrix 
$M_{\rm n,m}(\mathbf{k})$. The latter matrix, given in Eq.~\eqref{eq:M}, depends on $a_{\rm n}$, is symmetric, and 
contains all the scattering processes between fermions and photons plus the free fermion propagators on its diagonal. 
\begin{align}
&M_{\rm n,m}(\mathbf{k})=\nonumber\\
&\left(
\begin{array}{ccccccccccc}
 G_\psi^{-1}[0,0]   & 0 & 0 & 0 & 0 & \Lambda_{\rm n,m}/4  & \Lambda_{\rm n,m}/4  & \Lambda_{\rm n,m}/4  & \Lambda_{\rm n,m}/4  & U_{\rm n,m}/2 &\dots \\
 0 & G_\psi^{-1}[-1,0]   & U_{\rm n,m}/2 & \Lambda_{\rm n,m}/4  & \Lambda_{\rm n,m}/4  & 0 & 0 & 0 & 0 & 0 &  \\
 0 & U_{\rm n,m}/2 & G_\psi^{-1}[1,0]   & \Lambda_{\rm n,m}/4  & \Lambda_{\rm n,m}/4  & 0 & 0 & 0 & 0 & 0 & \\
 0 & \Lambda_{\rm n,m}/4  & \Lambda_{\rm n,m}/4  & G_\psi^{-1}[0,-1]   & U_{\Omega}/2 & 0 & 0 & 0 & 0 & 0 &   \\
 0 & \Lambda_{\rm n,m}/4  & \Lambda_{\rm n,m}/4  & U_{\Omega}/2 & G_\psi^{-1}[0,1]   & 0 & 0 & 0 & 0 & 0 &   \\
 \Lambda_{\rm n,m}/4  & 0 & 0 & 0 & 0 & G_\psi^{-1}[-1,-1] & U_{\Omega}/2 & U_{\rm n,m}/2 & 0 & \Lambda_{\rm n,m}/4  &   \\
 \Lambda_{\rm n,m}/4  & 0 & 0 & 0 & 0 & U_{\Omega}/2 & G_\psi^{-1}[-1,1]   & 0 & U_{\rm n,m}/2 & \Lambda_{\rm n,m}/4  &   \\
 \Lambda_{\rm n,m}/4  & 0 & 0 & 0 & 0 & U_{\rm n,m}/2 & 0 & G_\psi^{-1}[1,-1]  & U_{\Omega}/2 & 0 &   \\
 \Lambda_{\rm n,m}/4  & 0 & 0 & 0 & 0 & 0 & U_{\rm n,m}/2 & U_{\Omega}/2 & G_\psi^{-1}[1,1]   & 0 &    \\
U_{\rm n,m}/2 & 0 & 0 & 0 & 0 & \Lambda_{\rm n,m}/4  & \Lambda_{\rm n,m}/4  & 0 & 0 & G_\psi^{-1}[-2,0]   &   \\
 \vdots &&&&&& &&&&\ddots
\end{array}
\right)\;,
\label{eq:M}
\end{align}
where $\Lambda_{\rm n,m}=(g_0\Omega/\Delta_{\rm a})(a_{\rm
  m-n}^*+a_{\rm n-m}^{\phantom{*}})$, $U_{\rm n,m}=(g_0^2/2\beta\Delta_{\rm
  a})\sum_{\rm n_1}a_{\rm n1-n}^*a_{\rm n1-m}^{\phantom{*}}$, and $U_{\Omega}=\beta\Omega^2/2\Delta_{\rm a}$.
The diagonal entries are the fermion propagators  $G_{\psi}^{-1}[{\rm i,j}] =\beta\delta_{\rm n,m}(- i \nu_{\rm n} +
\xi_{\mathbf{k}}[{\rm i,j}]) + U_{\rm n,m} + U_{\Omega}$, with
\begin{align}
\xi_{\mathbf{k}}[{\rm i,j}] =
\frac{(k_x + {\rm i} Q_x)^2 + (k_y + {\rm j} Q_y)^2}{2m}-\mu_\psi\;.
\end{align}
We indicate bosonic and fermionic Matsubara frequencies by
$\omega_{\rm n}=\pi 2 {\rm n}/\beta$ and $\nu_{\rm n}=\pi (2 {\rm
  n}+1)/\beta$, respectively. In our numerical computations, this matrix 
  is truncated well in the regime where results converge.

The phase diagram can be obtained by a
saddle-point analysis of the action (\ref{S_eff}) at fixed density
$n_{\psi}=N/L^d$, as done in \cite{piazza13}. The corresponding mean-field
solution $a_{\rm n}=\beta\delta_{\rm n,0}\sqrt{N}\alpha$ becomes exact
in the thermodynamic limit $N,L\to\infty, n_{\psi}=\text{const.}$, where $\alpha\neq 0$ in the superradiant phase so that the number of cavity photons becomes finite.

\subsection{Dual effective action for fermionic atoms}
We here derive the dual effective action for the ground state fermionic atoms. 
After elimination of the excited, internal state the action corresponding to the Hamiltonian 
in Eq.~(1) of the main text reads
\begin{align}
S_{\psi\psi} &=
T\sum_n
\int d\mathbf{r}\,
\bar{\psi}_{\bf{r}}(\omega) 
\left\{
-i\omega_n
-\frac{\nabla^2}{2m}
+
\frac{\Omega^2}{\Delta_a}
-\mu_{\psi}
\right\}
\psi_{\bf{r}}(\omega)
\nonumber\\
S_{aa}&= \int_0^\beta d \tau\,
a^\ast(\tau) \frac{d a(\tau) }{d\tau} -\Delta_c a^\ast(\tau) a(\tau)
\nonumber\\
S_{\psi a}&= 
\int_0^\beta d\tau 
\int d\mathbf{r} \,
\bar{\psi}_{\mathbf{r}}(\tau)
\psi_{\mathbf{r}}(\tau)
\Bigg\{\frac{\left(g_0 \eta_{\mathbf{r}}\right)^2}{\Delta_a}a^\ast(\tau) a(\tau) 
+\frac{ \Omega g_0\eta_{\mathbf{r}}}{\Delta_a} 
\left[a(\tau) + a^\ast(\tau) \right]
\Bigg\} 
\;,
\end{align}
where we have used $\eta_{\rm p}(\rb)=1$, i.e. we here neglect the spatial structure of 
the pump laser and we dropped the cavity subscript on the cavity mode profile 
$\eta_{\mathbf{r}}$. We integrate out the photons and get
\begin{align}
S[\bar{\psi},\psi]= T\sum_{n,\mathbf{k}}\,
\bar{\psi}_{\bf{k}}(\omega) 
\left\{
-i\omega_n
+\xi_{\mathbf{k}}
\right\}
\psi_{\bf{k}}(\omega)
-
\frac{1}{2} \int_0^\beta d\tau  \int_0^\beta d\tau'
\int d\mathbf{r} \int d\mathbf{r}' 
n^{(\psi)}_{\bf{r}}(\tau) 
\left(
\frac{\Omega g_0}{\Delta_a} \eta_{\mathbf{r}}
\right)
\mathcal{G}(\tau-\tau')
\left(
\frac{\Omega g_0}{\Delta_a} \eta_{\mathbf{r}'}
\right)
n^{(\psi)}_{\bf{r}'}(\tau') 
\;.
\label{eq:S_ferm}
\end{align}
where the fermionic density is $n^{(\psi)}_{\bf{r}}(\tau)=\bar{\psi}_{\mathbf{r}}(\tau)
\psi_{\mathbf{r}}(\tau)$, the fermion dispersion is $\xi_{\mathbf{k}}=\frac{\bf{k}^2}{2m}-\mu$ and as before the 
standing cavity mode functions given by $\eta_{\mathbf{r}}=\cos(\mathbf{Q}_0\cdot \mathbf{r})$. 
The photon kernel mediating the interaction
\begin{align}
\mathcal{G}(\tau-\tau')= T\sum_{n} 
\frac{2 \mathcal{N}^{(\psi)}(\Omega) -\Delta_c}
{-\Omega_n^2 + \left( \mathcal{N}^{(\psi)}(\Omega) -\Delta_c \right)^2
}
e^{-i\Omega_n(\tau-\tau')}\;,
\end{align}
involves (weighted) spatial averages of the fermionic density fluctuations
\begin{align}
 \mathcal{N}^{(\psi)}(\Omega)=\int_{0}^\beta d\tau  \int d \mathbf{r}\, n^{(\psi)}_{\mathbf{r}}(\tau)
 \frac{\left( g_0 \eta_{\mathbf{r}} \right)^2}{\Delta_a}
 e^{i\Omega_n \tau}\;.
\end{align}
In the limit $\Delta_c\gg  \mathcal{N}^{(\psi)}(\Omega)$, the interaction becomes 
approximately instantaneous in (imaginary) time but remains long-ranged in space. 
Going now to momentum space for the fields
%
$\psi_{\mathbf{r}}(\tau)=\sum_{\mathbf{k}} \psi_{\mathbf{k}}(\tau) e^{-i \mathbf{k}\mathbf{r}}$,
$\bar{\psi}_{\mathbf{r}}(\tau)=\sum_{\mathbf{k}}\bar{\psi}_{\mathbf{k}}(\tau) e^{i \mathbf{k}\mathbf{r}}$
%
we obtain for the interaction term in Eq.~(\ref{eq:S_ferm})
\begin{align}
S[\bar{\psi},\psi]&= T\sum_{n,\mathbf{k}}\,
\bar{\psi}_{\bf{k}}(\omega) 
\left\{
-i\omega_n
+\xi_{\mathbf{k}}
\right\}
\psi_{\bf{k}}(\omega)\nonumber\\
&-
\frac{V_0}{N}
\int_0^\beta d\tau \sum_{\mathbf{k}_1,\mathbf{k}_2}
\frac{1}{2}\Big[ 
\bar{\psi}(\tau)_{\mathbf{k}_1-\mathbf{Q}_0}\psi_{\mathbf{k}_1}(\tau)
+
\bar{\psi}(\tau)_{\mathbf{k}_1+\mathbf{Q}_0}\psi_{\mathbf{k}_1}(\tau)
\Big]
\frac{1}{2}\Big[
\bar{\psi}(\tau)_{\mathbf{k}_2-\mathbf{Q}_0}\psi_{\mathbf{k}_2}(\tau)
+
\bar{\psi}(\tau)_{\mathbf{k}_2+\mathbf{Q}_0}\psi_{\mathbf{k}_2}(\tau)
\Big]
\;.
\end{align}
with
%
$\frac{V_0}{N} \equiv \left( \frac{\Omega g_0}{\Delta_a}\right)^2 \frac{1}{\Delta_c}$.
%
This action is now a reduced mean-field model for finite-$\mathbf{Q}_0$ 
charge-ordering of the Fermi liquid. Applying the same logic as in Ref.~\onlinecite{piazza13}, 
it can be shown that this model is exactly solvable in the thermodynamic limit $N\rightarrow \infty$.
The derivation of the free energy density proceeds in a standard way. We first 
decouple the interaction term with a Hubbard-Stratonovich field conjugate to the 
fermion bilinear
\begin{align}
\rho_{\mathbf{Q}_0}(\Omega)\leftrightarrow
T\sum_{n,\mathbf{k}} \frac{1}{2}
\Big[ 
\bar{\psi}_{\mathbf{k}-\mathbf{Q}_0}(\Omega+\omega)\psi_{\mathbf{k}}(\omega)
+
\bar{\psi}_{\mathbf{k}+\mathbf{Q}_0(\Omega+\omega)}\psi_{\mathbf{k}}(\omega)
\Big]
\end{align}
and we can in the following restrict to the static component of the order parameter field 
$\rho_{\mathbf{Q}_0}(\Omega)\rightarrow \frac{\delta_{\Omega,0}}{T} \rho$.
We now integrate out the fermions and obtain the free energy density
\begin{align}
\frac{F[\rho]}{N}
=
\frac{\rho^2}{V_0} - T\sum_n 
\int_{\mathcal{B}} \frac{d \mathbf{k}}{(2\pi)^d }
\text{tr}\ln
\left[
\mathcal{M}_\rho(\omega_n;\mathbf{k})
\right]
\label{eq:free_energy}
\end{align}
where the fermionic Nambu matrix can be written in the form
\begin{align} 
\mathcal{M}_{\alpha}(\omega_{\rm
  n};\kb)=
\left( 
\begin{array}{ccccccc}
\ddots & \vdots & \vdots & \vdots & \vdots & \vdots & \iddots  \\
0 & -\frac{\rho}{2} &
-i\omega_n +\xi_{\mathbf{k}-\mathbf{Q}_0} &
-\frac{\rho}{2}   & 0 &
 0 & 0\\
\dots& 0 &-\frac{\rho}{2}  &
  -i\omega_n +\xi_{\mathbf{k}}  &
-\frac{\rho}{2}  & 0 &
 \dots \\
0 & 0 & 0 & -\frac{\rho}{2}&
  -i\omega_n +\xi_{\mathbf{k}+\mathbf{Q}_0}  &
-\frac{\rho}{2} &  0 \\
 \iddots &\vdots&\vdots&\vdots&\vdots&\vdots&\ddots
\end{array} 
\right)
\label{eq:matrix}
\end{align}
By comparing Eq.~(\ref{eq:free_energy}) to the photon-only action Eq.~\eqref{S_eff}
we see that charge order parameter $\rho$ is dual to the cavity condensate $\alpha$.

\subsection{Bragg planes of the chequerboard lattice}

The density wave in the superradiant phase has
momentum $\mathbf{Q}=\mathbf{Q}_c+\mathbf{Q}_p$, corresponding to a
chequerboard lattice with reciprocal vector $\mathbf{Q}$ whose length
is $Q=\sqrt{Q_x^2+Q_y^2}$, as depicted in Fig.~\ref{fig:bragg}. This
standard textbook construction for a regular square lattice
\cite{ashcroft_mermin} allows to understand why a reconstructed Fermi
surface appears when $k_{\text F}>Q/2$. In the normal phase, when $k_{\text F}>Q/2$, the Fermi
surface (assumed circular to simplify the present discussion) crosses the first Bragg plane (more precisely a Bragg line in $d=2$). 
In this case, the Fermi
surface is split between the first and the second Brillouin zone,
creating empty corners in the former and filled arcs in the
latter. Slightly after entering the superradiant phase
where the chequerboard lattice is weak, small gaps appear at the Bragg
planes and the empty corners get separated from the filled arcs to
form the ``electron and hole pockets'' discussed in the text and
depicted in the upper right inset of Fig. 5 of the main text. The
repeated-zone representation used in the latter can be obtained from
the construction of Fig.~\ref{fig:bragg} by i) rejoining the different
segments of the each Brillouin zone to from a single whole square
Brillouin zone, ii) join equal copies of the whole zone to form the
periodic structure. 
There is, however, a difference between the repeated zone scheme
obtained from the above prescription and the repeated zone scheme used
in Fig. 5 of the main text. In the latter, the reference Brillouin
zone is $\mathcal{B}$ which is a square with side length $Q_x=Q/\sqrt{2}$. On
the other hand, in Fig.~\ref{fig:bragg} the Brillouin zone is a square
with side length $Q$.

\begin{figure}[h]	
  \includegraphics[width=80mm]{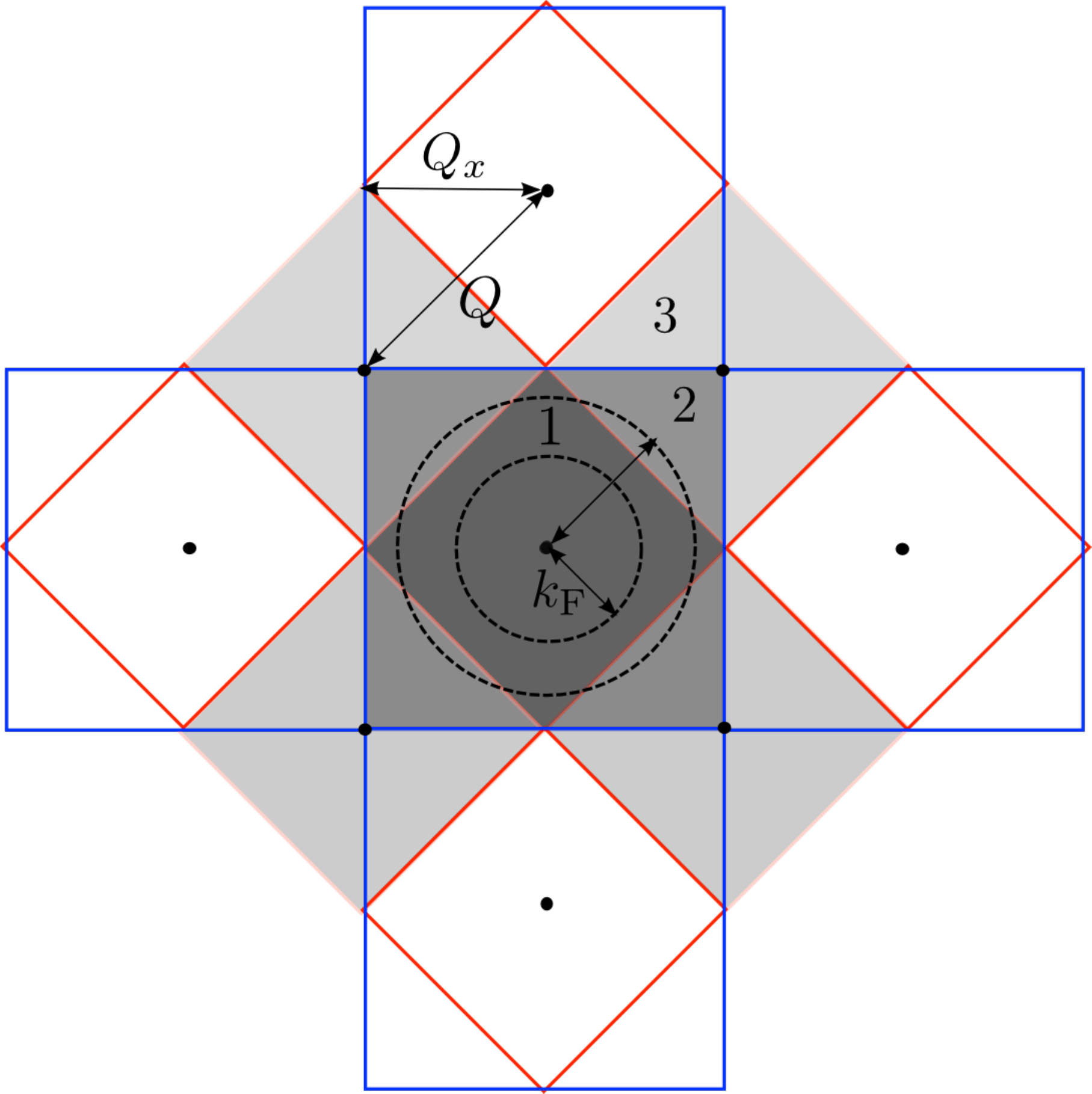}
\caption{
The reciprocal chequerboard lattice is a regular square lattice with
spacing $Q$ separating the lattice points (black circles). The blue
and red solid lines correspond to the first two Bragg lines while the
shaded areas represent the first three Brillouin zones in different
gray scales.
The two black-dashed circles show the Fermi surface for two different
values of $k_{\text F}$, the larger of which intersects the first
Bragg line. Here we took $Q_x=Q_y$ and assumed an circular Fermi
surface for simplicity.
}
\label{fig:bragg}
\end{figure}

\subsection{Cavity spectrum $A(\omega)$ and density response of the Fermi gas $\Pi_F(\omega,\mathbf{Q})$}

In order to obtain the cavity photon spectrum we expand the effective action \eqref{S_eff} about the saddle-point solution $\alpha$ up to second order, in the same way described in \cite{piazza13}. Out of the superradiant phase we have $\alpha=0$ and thus we can expand the effective action directly in $a_{\rm n}$ up to second order. This yelds the photon propagator from which we can compute the spectrum by analytical continuation, yielding the retarded propagator, and then by taking twice the imaginary part of the latter \cite{piazza13}.

The cavity photon spectrum, Eq.~(2) of the main text, 
can be expressed in completely analytic form in terms of the density response 
of the Fermi gas evaluated at fixed external momentum $\mathbf{Q}$
\begin{align}
\Pi_F(\omega_n,\mathbf{Q}) &= \int \frac{ d\omega'_n}{2\pi}
\int \frac{ d^d \mathbf{k}}{(2\pi)^d}
G_F(\omega'_n+\omega_n,\mathbf{k}+\mathbf{Q}) G_F(\omega'_n,\mathbf{k})
\nonumber\\
&=
\int \frac{ d\omega'_n}{2\pi}
\int \frac{ d^d \mathbf{k}}{(2\pi)^d}
\frac{1}{-i\left(\omega'_n+\omega_n\right) + \xi_{\mathbf{k}+\mathbf{Q}}}
\frac{1}{-i\omega_n + \xi_{\mathbf{k}}}
\nonumber\\
&=
\int \frac{ d^d \mathbf{k}}{(2\pi)^d}
\frac{n_{\rm{F}}(\xi_{\mathbf{k}+\mathbf{Q}}) - n_{\rm{F}}(\xi_{\mathbf{k}})}
{\xi_{\mathbf{k}}-\xi_{\mathbf{k}+\mathbf{Q}} - i\omega_n}\;,
\label{eq:def_Pi}
\end{align}
where the Fermi-Dirac distribution function $n_{\rm F}(x) = \frac{1}{\exp[x/T]+1}$ becomes the standard 
$\theta$-function at zero temperature $T=0$. We have added subscript $n$ to the frequency to remind ourselves 
that $\omega_n$ is an imaginary ``Matsubara'' frequency and that we have not yet analytically continued to real frequencies 
$\omega$. Note also that Eq.~(\ref{eq:def_Pi}) is defined with a relative minus sign with respect to the one-loop Feynman diagram (where an additional minus-sign appears). For rotationally invariant Fermi surfaces, $\xi_{\mathbf{k}} = \frac{\mathbf{k}^2}{2m}-\mu$, a number of analytical expressions are available for the real and imaginary part of $\Pi_{\rm F}(\omega,\mathbf{Q})$ 
after analytic continuation to real frequencies $\omega$ \cite{fukuyama91,bares93,metzner97,mihaila11}.

We give here only the 1d-expressions used to evaluate the cavity photon spectral function Eqs.~(2,3) and 
Figs. 2,3 of the main text, following the conventions of Ref.~\onlinecite{mihaila11} (multiplied by a global minus sign and adapted to spinless fermions).
The real part then is
\begin{align}
{\rm Re}\Pi_{\rm F,1d}(\omega,Q) = -\frac{n }{\epsilon_{\rm F} 4 \tilde{Q}} 
\left(
\log
\left |
\frac{1 + Q_{-}^2/(2 \tilde{Q})}{1 - Q_{-}^2/(2 \tilde{Q})}
\right |
-
\log\left |
\frac{1 + Q_{+}^2/(2 \tilde{Q})}{1 - Q_{+}^2/(2 \tilde{Q})}
\right |
\right)\;,
\label{eq:real_part}
\end{align}
with the abbreviations $Q_{\pm}^2 = \tilde{\omega}\pm \tilde{Q}^2$ where
$\tilde{\omega}=\omega/\epsilon_{\text F}$ and $\tilde{Q} = Q/k_{\rm F}$.

The imaginary part is
\begin{align}
{\rm Im}\Pi_{\rm F,1d}(\omega,Q) = \frac{n\pi}{\epsilon_{\rm F} 4 \tilde{Q}}\;,
\end{align}
and it differs from zero only if $\tilde{Q}\geq 2$ and $\tilde{Q}^2 - 2 \tilde{Q} \leq \tilde{\omega} \leq \tilde{Q}^2 + 2 \tilde{Q}$, or if $0\leq\tilde{Q}< 2$ and $-(\tilde{Q}^2 - 2 \tilde{Q}) \leq \tilde{\omega} \leq \tilde{Q}^2 + 2 \tilde{Q}$.

\end{widetext}

\end{document}